\documentclass{article}

\PassOptionsToPackage{numbers, compress}{natbib}
\usepackage[final]{neurips_2023}
\usepackage{color}
\usepackage{multirow}
\usepackage{caption}
\usepackage{wrapfig}
\usepackage{subcaption}
\usepackage{amsmath,amssymb}
\usepackage{graphicx}
\usepackage[toc,page,header]{appendix}
\providecommand{\gz}[1]{\textcolor{black}{{#1}}} 

\providecommand{\zd}[1]{\textcolor{black}{{#1}}}
\providecommand{\mac}[1]{\textcolor{black}{{#1}}}
\providecommand{\yw}[1]{\textcolor{black}{{#1}}}

\usepackage[utf8]{inputenc} 
\usepackage[T1]{fontenc}    
\usepackage{hyperref}       
\usepackage{url}            
\usepackage{booktabs}       
\usepackage{amsfonts}       
\usepackage{nicefrac}       
\usepackage{microtype}      
\usepackage{xcolor}         

\title{EDMSound: Spectrogram Based Diffusion Models for Efficient and High-Quality Audio Synthesis}

\author{%
  Ge Zhu\thanks{This work was partially done during an internship at Ubisoft La Forge.}~~~~Yutong Wen \\
Department of ECE, University of Rochester\\
  \texttt{\{ge.zhu,yutong.wen\}@rochester.edu} \\
  \And
  Marc-André Carbonneau \\
  Ubisoft La Forge \\
  \texttt{marc-andre.carbonneau@ubisoft.com} \\
  \And
  Zhiyao Duan \\
  Department of ECE, University of Rochester\\
  \texttt{zhiyao.duan@rochester.edu} \\
}

\begin{document}

\maketitle

\begin{abstract}
\mac{Audio diffusion models can synthesize a wide variety of sounds. 
Existing models often operate on the latent domain with cascaded phase recovery modules to reconstruct waveform.  
This poses challenges when generating high-fidelity audio.}
\gz{In this paper, we propose EDMSound, a diffusion-based generative model in spectrogram domain under the framework of elucidated diffusion models (EDM).}
Combining with efficient deterministic sampler, we achieve similar Fréchet audio distance (FAD) score as top-ranked baselines with only 10 steps and reach state-of-the-art performance with 50 steps on the DCASE2023 foley sound generation benchmark.
We also reveal a potential concern regarding diffusion based audio generation models that they tend to generate samples with high perceptual similarity to the data from training set. Project page: \url{https://agentcooper2002.github.io/EDMSound/}
\end{abstract}

\section{Introduction}
Audio synthesis research has a long history~\cite{moffat2019sound}. With the development of deep generative models in recent years, data-driven audio synthesis methods have become more and more popular. 
In particular, diffusion models~\cite{sohl2015deep,ho2020denoising} have led to transformative changes in audio synthesis tasks, resulting in higher quality audio samples.
\gz{Current diffusion-based audio generation models utilize a cascaded system design~\cite{liu2023audioldm,ghosal2023text,huang2023make,wang2023audit,huang2023noise2music,hawthorne2022multi} to circumvent the complexity of generating sound in the temporal domain~\cite{garbacea2019low}.}
\gz{They typically involve converting waveforms into spectrograms to train a base diffusion generator.} 
A secondary phase recovery network then converts the spectral representation back into the temporal domain~\cite{huang2023noise2music,hawthorne2022multi}.
To further reduce the computational complexity for the base diffusion model~\cite{liu2023audioldm,yang2023diffsound}, a variational autoencoder (VAE) can be used to transform a mel-spectrogram into a lower-dimensional latent representation.
However, a recent survey study~\cite{oh2023demand} suggests that current audio generation models might not be ready for professional sound production and \zd{the most significant challenge is presented in audio quality \gz{(e.g., fidelity, sampling rate, and level of noise)}}.
This audio fidelity degradation may be caused by the cumulative errors across modules in cascaded frameworks~\cite{pascual2023full}.
\gz{In addition, existing diffusion-based audio generation systems are inefficient at inference time and vanilla samplers will typically take hundreds of neural function evaluations (NFEs). For instance, AudioLDM~\cite{liu2023audioldm} takes 100-200 NFEs with denoising diffusion implicit model (DDIM)\cite{song2020denoising} for better sample quality.}

\mac{In this study, we target at improving the generation fidelity in an end-to-end manner by developing an audio generation model in the complex spectrogram domain.}
\gz{Compared to magnitude spectrum and phase spectrum, the real and imaginary components of the complex spectrograms exhibit clear structures and are suitable for deep learning models~\cite{richter2023speech}.}
\gz{Compared to raw waveform modeling~\cite{pascual2023full}, spectral features have fewer temporal redundancies~\cite{liu2023simple}.}
To improve generation fidelity, we build our diffusion generators within the framework of EDM~\cite{karras2022elucidating} due to its SoTA performance in several image generation benchmarks.
To accelerate the generation while maintaining similar sample quality, we use exponential integrator (EI) based ordinary differential equation (ODE) samplers during inference~\cite{zhang2022fast,lu2022dpm,lu2022dpmpp}.
We validate our method on different sound categories using DCASE2023 foley sound generation benchmark and Speech Command 09 (SC09)~\cite{warden2018speech} dataset (containing spoken digits from `zero' to `nine') using Fréchet distance as evaluation metric for its correlation with human perception~\cite{kilgour2018fr}.

While diffusion-based models are capable of generating high quality audio samples, it can unintentionally replicate training data~\cite{somepalli2022diffusion}. 
\yw{Replicating data might also harm the audio generation diversity.}
Although similar concerns have been explored in computer vision by~\cite{somepalli2022diffusion, somepalli2023understanding}, examination of this issue in audio generation remains an open research area. 
\yw{In our work, we answer} the question of whether diffusion-based models generate audio samples with replicated contents.

To summarize, we introduce an end-to-end audio diffusion model, EDMSound, in the complex spectrogram domain.
\yw{At inference time, we use EI-based ODE samplers to accelerate the generation speed.}
\gz{We achieve the SoTA performance on DCASE2023 foley sound generation benchmark and competitive performance on SC09 dataset in terms of Fréchet distance.}
We propose a method to examine the memorization issue, \textit{i.e.,} content replication on a range of diffusion-based audio generation models on the DCASE2023 benchmark dataset. 
Qualitative and quantitative analysis show that our proposed model does not generate exact copies of training data. 
Instead, it is able to generate audio samples that match the sound timbre of the training samples but with varied temporal patterns.



\section{Method}
\label{sec:method}

\vspace{-1cm}
\begin{center}
    \begin{table*}[!t]
\centering
\caption{FAD score and relative dataset similarity score comparison of the generated audio samples on DCASE2023 task7. 
Baseline systems that rank 1st in the challenge, both achieved top3 official rank.
`mean' represents the average value of the experiments.
`best' represents the best one.}
\tiny
\setlength\tabcolsep{3pt}
\begin{tabular}[width=\linewidth]{lcccccccc}
\toprule
\multirow{2}{*}{System}    & Dog   & \multirow{2}{*}{Footstep} & \multirow{2}{*}{Gunshot}    & \multirow{2}{*}{Keyboard}     &Moving Motor  & \multirow{2}{*}{Rain}          & Sneeze & \multirow{2}{*}{Overall}\\
&Bark&&&& Vehicle & &Cough& \\
\toprule
\textit{FAD score}~($\downarrow$)\\
Scheibler \textit{et al}~\cite{scheibler2023class}& 3.68	&8.07	&3.65	&\textbf{2.78}	&\textbf{7.42}	&5.23	&2.61& 4.78\\
Yi \textit{et al}~\cite{yuan2023latent}&3.62&5.10	&5.74	&3.04&9.80	&5.96	&1.90 &5.02 \\
Jung \textit{et al}~\cite{chung2023foley}&3.34	&3.99	&3.50	&4.07	&14.86	&\textbf{3.53}	&1.87& 5.02\\
\textbf{EDMSound-mean (Ours)}&\textbf{2.93}&\textbf{3.22}&\textbf{3.61}&3.73&11.10&6.01&\textbf{1.27}&\textbf{4.56}\\
\midrule
\textit{Relative dataset similarity}\\
Scheibler \textit{et al}~\cite{scheibler2023class}&\textbf{-0.02}	&\textbf{-0.04}	&\textbf{-0.04}	&\textbf{-0.07}	&\textbf{-0.02}	&-0.09	&\textcolor{red}{\textbf{0.03}} & \textbf{-0.04}\\
Yi \textit{et al}~\cite{yuan2023latent}&-0.05	&-0.07	&-0.11&-0.08	&-0.03	&-0.04 &-0.05 &-0.06 \\
Jung \textit{et al} (closed)~\cite{chung2023foley}&-0.14	&-0.11	&-0.11	&-0.18	&-0.10	&-0.17	&-0.11& -0.13\\
\textbf{EDMSound-best (Ours)}&-0.05&-0.06&-0.06&-0.08&-0.02&\textbf{-0.02}&-0.11&-0.05\\
\bottomrule
\end{tabular}
\label{table:results}
\vspace{-2mm}
\end{table*}

\end{center}

Diffusion probabilistic models (DPMs)~\cite{ho2020denoising,sohl2015deep} 
\gz{involve (1) corrupting training data with increasing noise levels into normal distribution and (2) learning to reverse each step of this noise corruption with the same functional form.}  
It can be generalized into score-based generative models~\cite{song2020score} which employ an infinite number of noise scales so that both forward and backward diffusion processes can be described by stochastic differential equations (SDEs).
During inference, \gz{the reverse SDE} is used to generate samples with numerical approaches starting from the standard normal distribution.
A remarkable property of the reverse SDE is the existence of a deterministic process, namely \textit{probability flow ODE}, whose trajectories share the same marginal probability as the original SDE~\cite{song2020score}.
As a result, one can employ ODE solvers. 
These solvers, in contrast to SDE solvers, allow for larger step sizes, primarily because they are not influenced by the inherent randomness of the SDE~\cite{lu2022dpm}.

\textbf{EDM on Complex Spectrogram} We train our diffusion model using EDM~\cite{karras2022elucidating} which formulates the above diffusion SDE with noise scales instead of drift and diffusion coefficients.
Practically, it presents a systematic way to design both training and sampling processes.
\gz{To ensure that the neural network inputs are scaled within $[-1,1]$ required by the diffusion models}, we apply an amplitude transformation on the complex spectrogram inputs, $\tilde c = \beta|c|^{\alpha}e^{i\angle{c}}$ following~\cite{richter2023speech}, where $\alpha\in (0, 1]$ is a compression factor which emphasize time-frequency bins with low energy, $\angle{c}$ represents the angle of the original complex spectrogram, and $\beta \in \mathbf{R}_+$ is a scaling factor to normalize amplitudes roughly to within $[0, 1]$. 
Such compression technique was originally proposed for speech enhancement~\cite{breithaupt2010analysis}, but we found it also effective in general sounds. 
We adopt 2D efficient U-Net proposed in Imagen~\cite{saharia2022photorealistic} as our diffusion model backbone due to its \gz{high} sample quality, faster convergence speed and memory-efficiency. 
During training, we use preconditioned denoising score matching as our training objective following~\cite{karras2022elucidating}. 
\textit{i.e}, $\mathbb{E}_{\mathbf{x}} \mathbb{E}_{\mathbf{n}} [\lambda(\sigma)\lVert D(\mathbf{x} + \mathbf{n}; \sigma) - \mathbf{x} \rVert^2_2]$, where $\mathbf{x}$ is the training data and $\mathbf{n}\in \mathcal{N}(\mathbf{0},\sigma^2\mathbf{I})$. 
We apply classifier free guidance (CFG)~\cite{ho2022classifierfree} at the sampling stage in the conditional generation task.


\begin{figure}[!t]
  \begin{minipage}[b]{0.49\textwidth}
    \centering    {\includegraphics[width=1\linewidth]{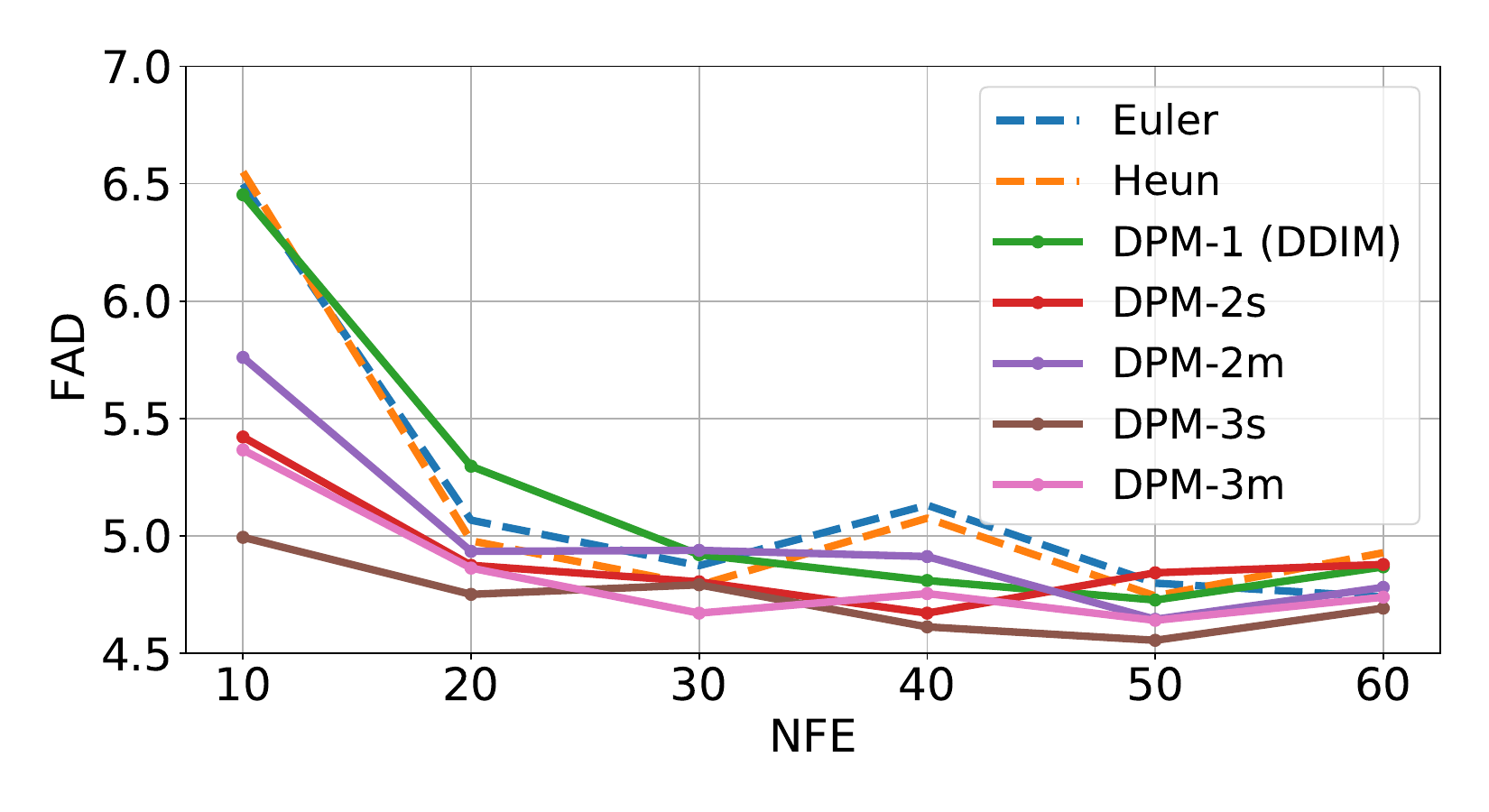}}
    \hspace{-1mm}
    \captionof{figure}{Comparison of FAD scores using different ODE samplers on DCASE 2023 Task 7. In DPM based samplers, the number indicates the order of the solvers, `s' represents `singlestep' and `m' represents `multistep'. We use CFG with a scale of 2.0 and repeat experiments three times.}
    \label{fig:samplers}
  \end{minipage}
  \hfill
  \begin{minipage}[b]{0.49\textwidth}
    \centering
    \tiny
    \begin{tabular}[0.3\linewidth]{llllllllllll}
            \toprule
            Model & Params & FID $\downarrow$ & IS $\uparrow$ & mIS $\uparrow$ & AM $\downarrow$\\
            \midrule
            \textit{Autoregressive}\\
            SampleRNN~\cite{mehri2016samplernn}& 35.0M  & 8.96  & 1.71  & 3.02    & 1.76  \\
            WaveNet~\cite{oord2016wavenet} & 4.2M & 5.08 & 2.27& $5.80$  & 1.47\\
            Sashimi~\cite{goel2022s} & 4.1M   & 1.99 & 4.12 & 24.57   & 0.90 \\
            \midrule
            \textit{Non-autoregressive}\\
            WaveGAN~\cite{donahue2018adversarial} & 19.1M   & 2.03  & 4.90 & 36.10 & 0.80 \\
            DiffWave~\cite{kong2020diffwave}  & 24.1M    & 1.92   & 5.26  & 51.21   & 0.68 \\
            ~~w/ Sashimi & 23.0M  & 1.42 & 5.94  & 69.17 & 0.59 \\
            ASGAN (Mel.)~\cite{baas2023gan} & 38.0M  &0.56&7.02 &\textbf{162.8} &0.56 \\
            ASGAN (HuBERT) & -  &\textbf{0.14} & \textbf{7.67} &\textbf{226.7}  &\textbf{0.26}\\
            EDMSound (Ours)  & 45.2M    & \textbf{0.14} & 7.17  & 160.2   & 0.33 \\
            \midrule
            Train & -   & 0.00   & 8.56  & 292.5 & 0.16 \\
            Test & -  & 0.02 & 8.33 & 257.6 & 0.19 \\
            \bottomrule
        \end{tabular}
        \captionof{table}{Comparison of unconditional generation with automated metrics on SC09 dataset. FID (Fréchet Inception Distance), IS (Inception score), modified IS, and AM score are measures for generated diversity and quality.}
        \label{tab:sc09}
    \end{minipage}
  \end{figure}

\textbf{Efficient Diffusion Samplers} 
Within EDM, the probability flow ODE can be simplified as a nonlinear ODE, enabling the direct application of standard off-the-shelf ODE solvers.
It is found that EI based ODE solvers have the minimum error with limited steps in a semilinear structured ODE~\cite{zhang2022fast}, a combination of a linear function of the data variable and a nonlinear neural network term.
Though such probability flow ODE only contains the non-linear neural network in EDM case, it is still beneficial to integrate EI approach shown in~\cite{zhang2022fast}.
\gz{Therefore, we use high order EI based ODE solvers~\cite{lu2022dpm}, \textit{i.e.,} singlestep and multistep DPM-solvers~\cite{lu2022dpm,lu2022dpmpp}.}

\textbf{Content Replication Detection} We define content replication as the presence of \gz{generated samples that are either \textit{complete duplicates} or a \textit{substantially similar portions} of the training samples.} 
\gz{It is found that representations trained with full supervision or self-supervised learning can perform as well as detectors specially trained for content replication detection~\cite{somepalli2022diffusion}.}
\gz{Since there is no existing content replication detectors for sound effects generation, we employ off-the-shelf pretrained audio representations including CLAP~\cite{laionclap2023}, AudioMAE~\cite{huang2022masked}, and PANNs~\cite{kong2020panns}, and compute the cosine similarity score to measure the degree of content replication.}
\gz{To better adapt the audio descriptors for this task, we conduct an additional fine-tuning stage:} We attach multi-layer perceptrons to the frozen pre-trained audio encoders and then train with triplet margin loss~\cite{schroff2015facenet}.
To enhance the robustness of the descriptor, we incorporate data augmentation by injecting Gaussian noise, random amplitude scaling, and temporal shifting to audio samples. 
\gz{We first choose one audio sample as an \textit{anchor sample} and a \textit{positive pair} with the same audio with the above augmentation.} 
\gz{Then, we randomly select another audio sample within the same class as the \textit{negative pair} with data augmentation}.
After the fine-tuning step, we search the training set based on the cosine similarity for each generated audio sample. 
We identify \textit{matched audio samples} within the training set with the top-1 similarity scores. 
These identified training samples are regarded as top matches for their corresponding generated audio counterparts. 
\yw{We then analyze the content replication behavior within these matched pairs.}
\vspace{-0.2cm}
\section{Experiment}
\vspace{-0.1cm}
\textbf{Experimental setup} 
We benchmark our model, EDMSound, on DCASE2023 task7 and SC09 dataset.
\gz{DCASE2023 foley sound generation~\cite{choi2023foley} aims at improving audio fidelity, fit-to-category, and diversity for foley sound generation and it provides a standardized evaluation framework for different systems.} 
It includes an open track and a closed track regarding the training dataset scale: the open track allows participants to leverage the datasets beyond the development set, while the closed track limits the dataset usage. 
We compare with strong baseline systems on both tracks in terms of FAD. 
Yi \textit{et al.}~\cite{yuan2023latent} (ranked 1st officially) and Scheibler \textit{et al.}~\cite{scheibler2023class} (achieved the highest FAD score on the open track) use LDMs. 
Jung \textit{et al.}~\cite{chung2023foley} use a GAN-based model, and ranked 1st in FAD score on the closed track. 
For the SC09 benchmark for unconditional generation, we retrain EDMSound without CFG following the best sampler choice and compare our result with baselines including autoregressive models as well as non-autoregressive models.

\textbf{Sound Generation Evaluation} We present our average FAD with a comparative assessment against baseline models on DCASE2023 foley sound dataset. 
We first compare generic ODE solvers Euler and Heun with EI-based DPM-solvers shown in Fig.~\ref{fig:samplers}.
It can be seen that the higher order EI-based ODE solvers yield faster convergence, \gz{therefore fewer NFEs are required during inference}.
Particularly, the 3rd order single-step DPM-solver (DPM-3s) reaches the similar FAD as Yi \textit{et al.}~\cite{yuan2023latent} with only 10 steps and achieves the best score 4.56 at the 50th step.
Tab.~\ref{table:results} presents the class-wise and overall FAD scores for the DACSE2023 task7 challenge. 
Our proposed method, EDMSound, outperforms the baseline models in terms of the overall FAD in both open and closed tracks, exhibiting better performance across the majority of class-specific FAD scores as well.
In Tab.~\ref{tab:sc09}, we present the SC09 benchmark result. 
Particularly, we achieved the lowest FID score without pretrained speech representations in \cite{baas2023gan}. 
These results underline the efficacy of our proposed methodology across diverse evaluation benchmarks. 

\textbf{Copy Detection Evaluation} 
\mac{We evaluate whether the generative models produce samples copied from training dataset on the DCASE2023 task7 dataset~\cite{choi2023foley} which contains a wide variety of sound types. We compare four systems:} the EDMSound-best, Scheibler \textit{et al.}~\cite{scheibler2023class} and Jung \textit{et al.}~\cite{chung2023foley} with the FAD scores being 4.36, 4.78, and 5.02 respectively. 
For content replication detection, we report our results using the fine-tuned CLAP audio encoder based on two observations: First, the similarity score distribution is significantly broader when fine-tuned with CLAP, highlighting its capability in distinguishing similar samples. Secondly, it demonstrates close alignment with human perception, as verified through manual examination.
Further details regarding the ablation study on other audio descriptors can be found in Appendix~\ref{appb}.

\begin{figure}[t]
  \centering  \includegraphics[width=0.7\linewidth]{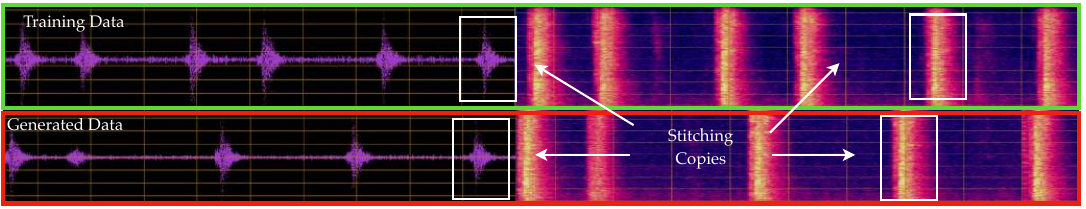}
  \caption{A matched pair of audio samples found by fine-tuned CLAP that shows a clear sign of stitching copy. The sound sources in the two audio samples show high similarity.} 
  \label{fig:copy}
\end{figure}

After mining the top-1 matched audio samples, we observe high resemblance between a number of pairs from our EDMSound-best.
Fig.~\ref{fig:copy} illustrates the waveform and spectrogram of a paired audio segment capturing the sound of keyboards. 
Despite variations in the temporal alignment of key presses, the spectral coherence strongly suggests a common sound source originating from the same keyboard. 
We thereby infer that our diffusion model imitates the characteristics of its training dataset. 
We term the phenomenon where generated samples closely match the training samples in spectral coherence, but with relaxed temporal alignment, as \textit{`stitching copies'}.
\mac{After listening to samples from all systems, we find that the model from Scheibler \textit{et al.} is prone to producing samples that are `exact copies' with training samples. 
This strong similarity probably suggests over-fitting in the task of generating foley sounds using a large, pre-trained LDM-based model with more than 850 million parameters~\cite{ghosal2023text}.}
For a more comprehensive visual representation of this phenomenon, please refer to our project page. 
\yw{To quantify the overall similarity between the generated data and the training data, 
we use the 95-percentile similarity score of all \textit{matched audio samples} defined in Sec.~\ref{sec:method}.}
To better compare the distribution difference, we compute the relative similarity by subtracting the training dataset similarity scores from the generated ones shown in the lower part of Tab.~\ref{table:results}. 
Despite the fact that there are instances from generated samples showing a high degree of similarity, the overall negative relative similarity scores indicate that none of the generative models replicate their training set more than the intrinsic similarity within the training set itself.


\vspace{-0.2cm}
\section{Conclusion}
\vspace{-0.2cm}
\mac{This paper introduced EDMSound, a simple and effective end-to-end diffusion model working on the complex spectral domain implementing efficient ODE solvers. 
EDMSound synthesizes high quality audio improving the state-of-the-art in terms of FAD on two standard benchmark datasets (SC09 and DCASE2023 challenge task7). 
Furthermore, we proposed fine-tuned CLAP to examine the issue of content replication in the audio domain. }


\section{Acknowledgement}
This work was partially supported by a Goergen Institute for Data Science (GIDS) Seed Funding Award at the University of Rochester.
{\small
  \bibliographystyle{ieee}
  \bibliography{main}
}
\newpage

\begin{appendices}
\section{Training and sampling within EDMs}
When using complex spectrograms as the diffusion model inputs, the real and imaginary components are treated as two separate channels corrupted by Gaussian noise in the forward process.
And as a result, the phase information is gradually destroyed. 
In the reverse sampling process, the real and imaginary channels are gradually recovered through the score network and thereby recover the phase.
\subsection{Training}
In DPMs, the neural networks are usually used to model the score~\cite{song2020score} of the data at each noise level, $\nabla_{\mathbf{x}}\log(\mathbf{x};\sigma)$, where $\mathbf{x}$ is the data and $\sigma$ is the noise level.
\textit{i.e.}, the the gradient of the log probability desnity with respect to data.
Or equivalently, it can be seen as training a denoiser function~\cite{karras2022elucidating} $D(\mathbf x;\sigma)$ to recover clean data given corrupted versions, where $\nabla_{\mathbf{x}}\log(\mathbf{x};\sigma)=(D (\bf{x};\sigma)-\mathbf{x})/\sigma^2$.
However, its magnitude varies significantly on given noise level. 
To decorrelate the magnitude of the network prediction with $\sigma$, we follow the preconditioning scales on the denoiser function proposed in~\cite{karras2022elucidating} with $c_{skip}(\sigma)\mathbf{x}+c_{out}(\sigma)F_{\theta}(c_{in}(\sigma)\mathbf{x};c_{noise}(\sigma))$,
where $F_{\theta}(\cdot)$ is the neural network output, $c_{in}$, $c_{out}$ are set to ensure a unit variance for the network inputs and outputs, $c_{skip}$ is set to minimize $F_{\theta}$ prediction errors scaled by $c_{out}$, and $\lambda(\sigma)$ is set to $1/c_{out}^2(\sigma)$ to equalize the initial training loss.
Following Karras et al.~\cite{karras2022elucidating}, the desnoiser preconditioning can be written as:
\begin{equation}
    D(\mathbf{x};\sigma)=\frac{\sigma_{data}^2}{\sigma_{data}^2+\sigma^2}\mathbf{x} + \frac{\sigma\cdot\sigma_{data}}{\sqrt{\sigma_{data}^2+\sigma^2}} F_\theta\Big(\frac{\mathbf{x}}{\sqrt{\sigma_{data}^2+\sigma^2}};\frac{\ln(\sigma)}{4}\Big).
\end{equation}
During training, we use $\sigma_{data}=0.2$ as the approximation for the standard deviation of the compressed input spectrogram magnitude values.
For $\sigma$, we use the log normal noise distribution with mean of -3.0 and variance of 1.0.
Notice that we did not tune these distribution parameters due to insufficient computation budgets, but we found that when synthesizing sounds without much semantic information, the final performance is robust to reasonable parameters.
Finally, we can write the training objective as:
\begin{equation}
    \mathbb{E}_{p_{data}(\mathbf{x}),\mathbf{\epsilon}, \sigma}[\lambda(\sigma)\Vert D(\mathbf{x} + \sigma \mathbf{\epsilon};\sigma) - \mathbf{x} \Vert^2_2],
\end{equation}
where $p_{data}(\mathbf{x})$ represents the training data distribution, $\mathbf{\epsilon}\sim\mathcal{N}(\mathbf{0}, \mathbf{I})$ is the standard normal distribution, $\sigma$ is the noise levels during training, and $\lambda(\sigma)$ is the loss weighting factor.

\subsection{Sampling}
Efficient samplers for DPMs can be categorized into training-based methods and training-free methods.
Of all the efficient samplers, training-free methods offer the advantage of direct applicability to pretrained DPMs without necessitating additional training phases~\cite{zhao2023unipc}.
Recently proposed fast training-free samplers are using ODE solvers since SDE solvers are hard to converge within a few steps due to the fact that discretizing SDEs is generally difficult in high dimensional space and is limited by the randomness of the Wiener process~\cite{lu2022dpm,kloeden1992stochastic}.
Another benefit of using ODE solvers is that such deterministic sampling is able to map the input data into corresponding latent representations and useful for editing.

In EDM, the probability flow ODE can be formulated as: 
\begin{equation}
\label{eq:ode}
\mathrm{d} \mathbf{x} = -\dot\sigma(t) ~\sigma(t) ~\nabla_{\hspace{-0.5mm}\mathbf{x}} \log p \big( \mathbf{x}; \sigma(t) \big) ~\mathrm{d} t \text{.}
\end{equation}
This simplification enables the direct application of standard off-the-shelf ODE solvers.
When numerically solving ODEs, each step introduces a local error, which cumulatively results in a global error over a specified number of steps.
The commonly used Euler’s method is a first order ODE solver with global error linearly proportional to step size. 
Higher order solvers have lower global error at a given time but require multiple NFEs at each step.
The second order Heun solver~\cite{karras2022elucidating} provides a good trade-off between global error and NFE. 
With the advantage of EI-based ODE solvers, we apply DPM solvers~\cite{lu2022dpmpp}, and our DPM-solvers samplers codebase are adapted from the implementation of k-diffusion~\footnote{https://github.com/crowsonkb/k-diffusion} and DPM-solver official codebase~\footnote{https://github.com/LuChengTHU/dpm-solver}.
During inference, we use Karras scheduler proposed in EDM with a $\rho$ of 7.0, $\sigma_{min}$ of 0.0001 and $\sigma_{max}$ 3.0.
We also a dynamic threshold of 0.99 following Imagen~\cite{saharia2022photorealistic}.
For conditional generation, we use CFG scale of 2.0 as we found it achieved the best performance across different samplers.

\subsection{Neural networks}
We applies efficien U-Net as our denoiser function backbone, which is designed to be memory efficient and converge fast.
It reverses the order of downsampling/upsampling operations in order to improve the speed of the forward pass.
For more detailed descriptions of the architecture, we encourage the readers to Appendix B from~\cite{saharia2022photorealistic}.
Our efficient U-Net is adapted from open source Imagen\footnote{https://github.com/lucidrains/imagen-pytorch}, 
For the input complex spectrogram, we use short-time Fourier transform (STFT) with window size of 510 samples and hop size of 256 samples.
We use an input channel of 2 for the real and imaginary components, 128 as the base dimension and channel multipliers of [1, 2, 2, 2].
For each downsampling/upsampling block, we use 2 ResNet blocks with 2 attention heads in self-attention layer. 
The model has a total of 45.2 million trainable parameters.
We use class label and $\log \sigma$ as efficient U-Net conditional inputs. 
For class conditioning, we represent class labels as one-hot encoded vectors, and then feed them through a fully-connected layer.

\section{Comparison of audio descriptors in copy detection}
\label{appb}
\begin{figure}[h]
\begin{subfigure}{0.48\textwidth}
\centering
\includegraphics[width=\linewidth]{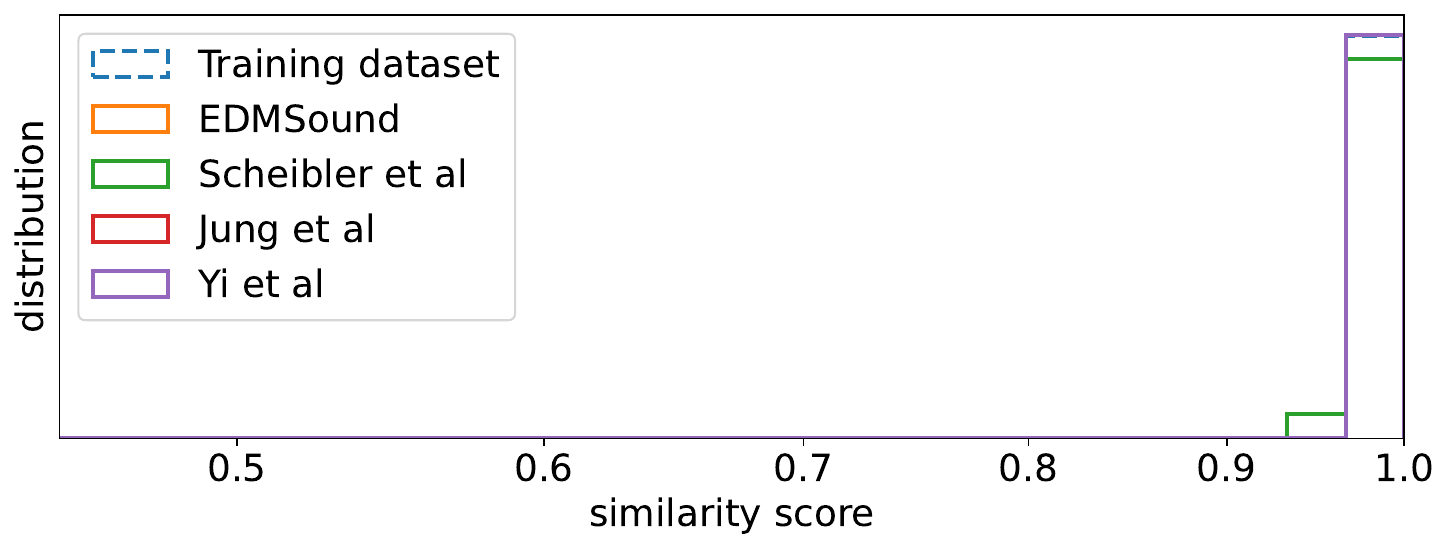}
  \caption{AudioMAE zero-shot.} 
  \label{fig:abcp_1}
\end{subfigure}
 \hfill
\begin{subfigure}{0.48\textwidth}
\centering
\includegraphics[width=\linewidth]{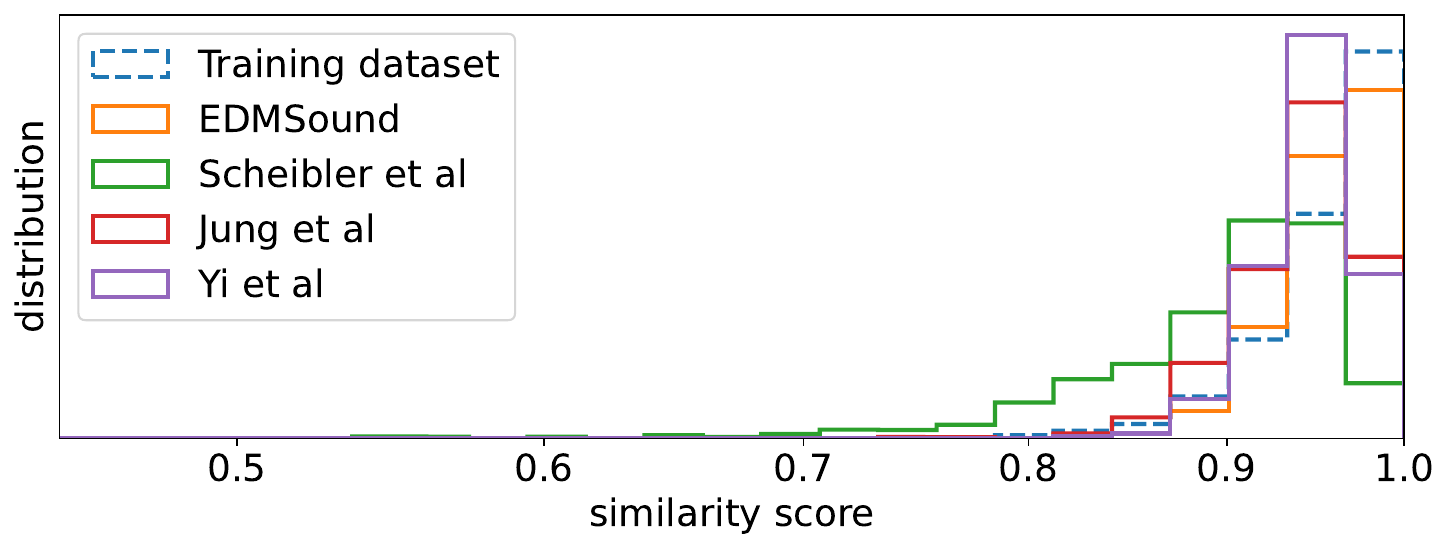}
  \caption{AudioMAE fine-tuned.} 
  \label{fig:abcp_2}
\end{subfigure}

\begin{subfigure}{0.48\textwidth}
\centering
\includegraphics[width=\linewidth]{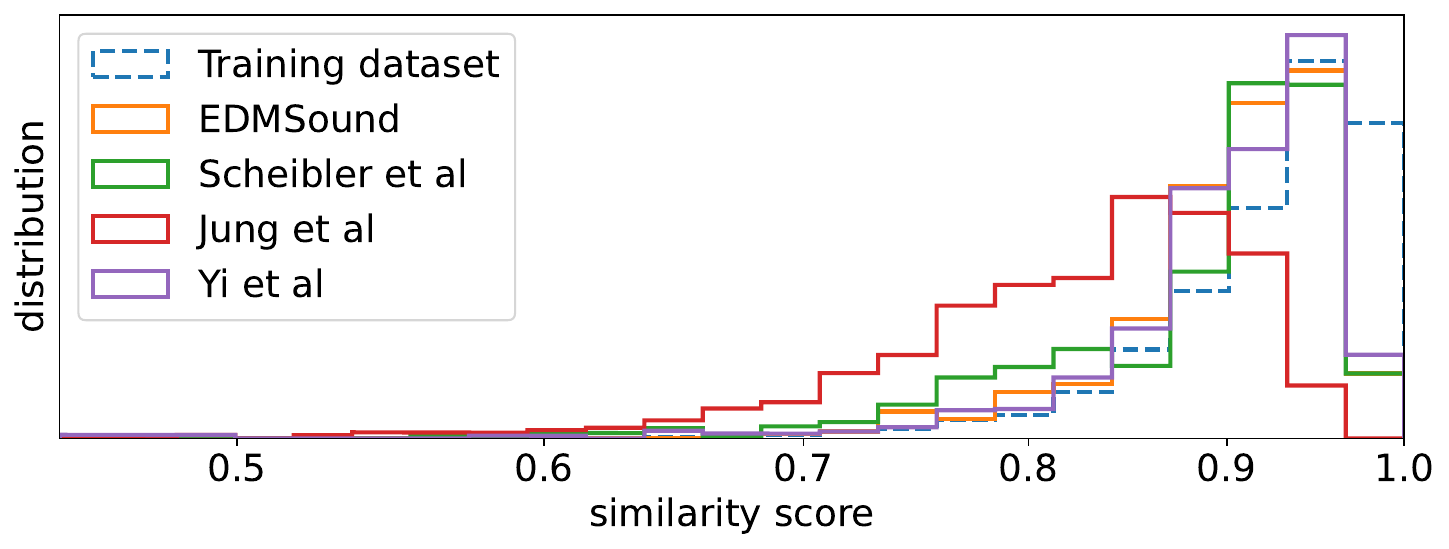}
  \caption{CLAP zero-shot.} 
  \label{fig:abcp_3}
\end{subfigure}
 \hfill
\begin{subfigure}{0.48\textwidth}
\centering
\includegraphics[width=\linewidth]{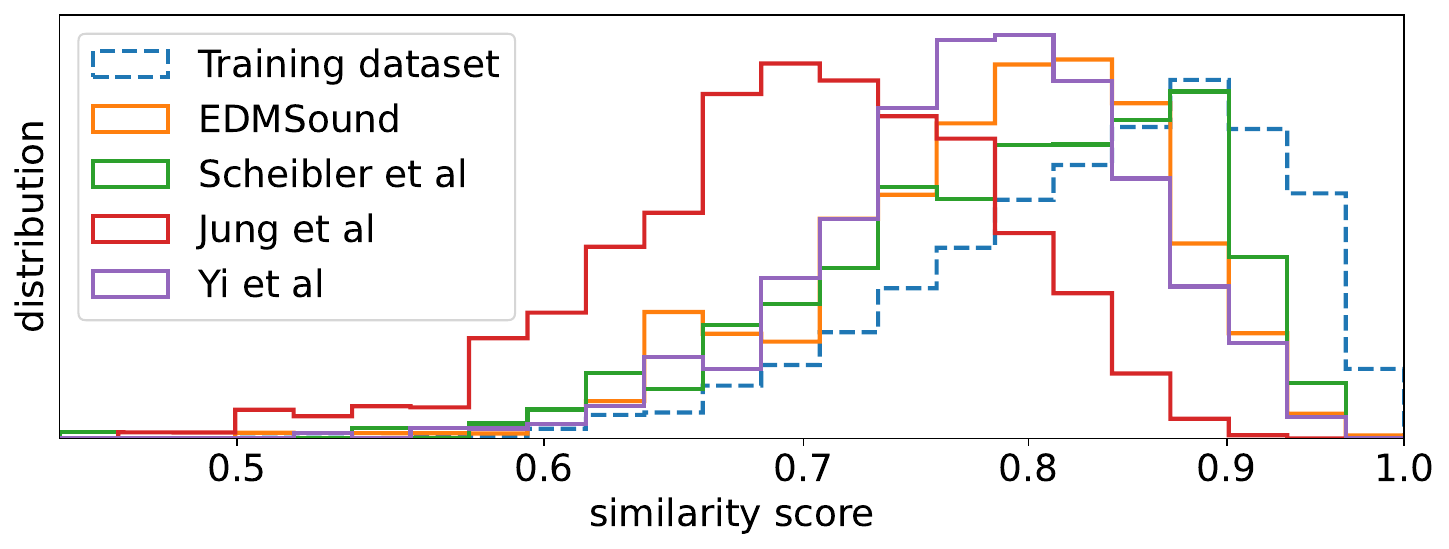}
  \caption{CLAP fine-tuned.} 
  \label{fig:abcp_4}
\end{subfigure}

\begin{subfigure}{0.48\textwidth}
\centering
\includegraphics[width=\linewidth]{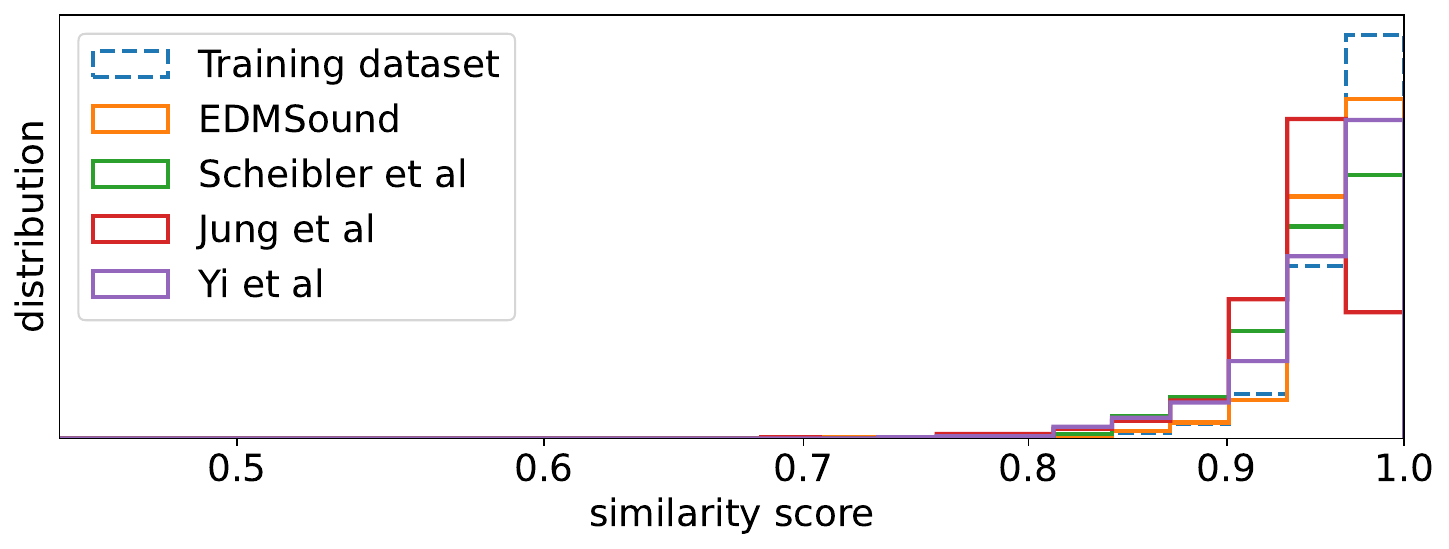}
  \caption{PANNs zero-shot.} 
  \label{fig:abcp_5}
\end{subfigure}
 \hfill
\begin{subfigure}{0.48\textwidth}
\centering
\includegraphics[width=\linewidth]{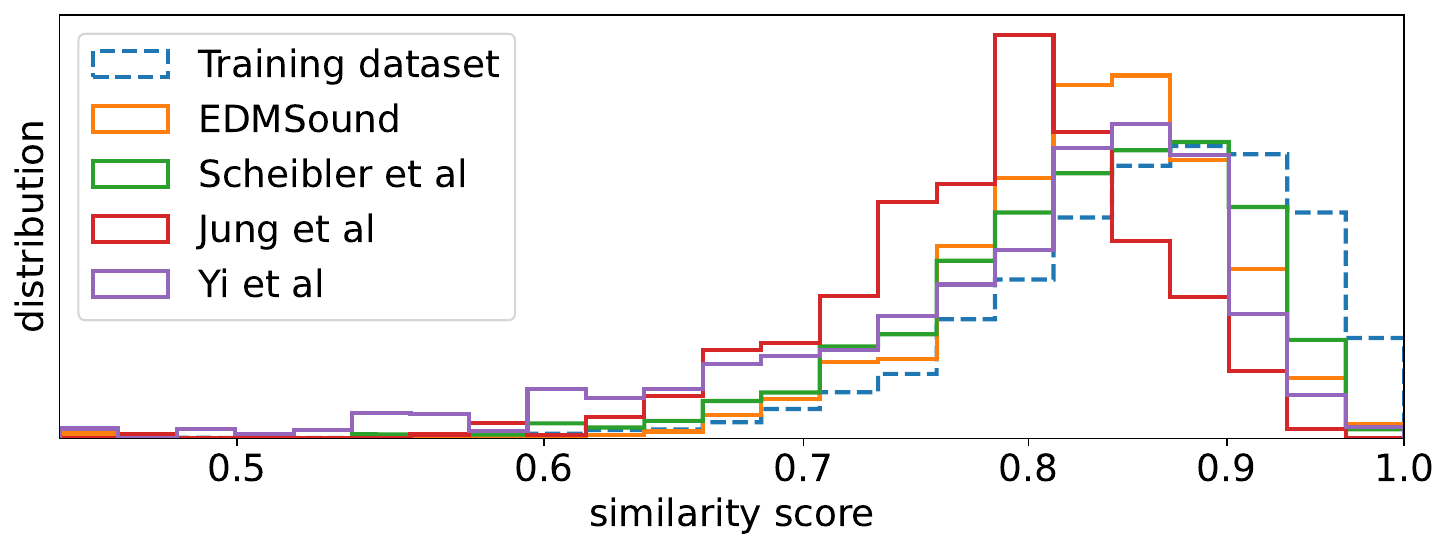}
  \caption{PANNs fine-tuned.} 
  \label{fig:abcp_6}
\end{subfigure}
\caption{Dataset similarity distribution of top1 matched pairs comparison of AudioMAE, CLAP and PANNs.}
\label{fig:descriptor_ab}
\end{figure} 

In this section, we compare the dataset similarity distributions from the DCASE2023 challenge task7 systems computed using the three audio encoder both with and without the fine-tune process. 
Fig.~\ref{fig:descriptor_ab} shows the comparison results. 
In the left column, we present the similarity distribution of audio encoders in a zero-shot copy detection scenario (\textit{i.e.}, without fine-tuning). 
The right column presents the outcomes post fine-tuning.
From the figure, we can observe that before fine-tuning, the similarity scores are close to 1 especially for AudioMAE, and this suggests that the audio representations are too close for intra-class samples. 
This fine-tuning helps to discriminate audio samples within the same class as the similarity score distribution masses shift to the left and spread out compared to the models without fine-tuning.
Upon evaluation, the fine-tuned CLAP model exhibits the most distinctive distribution spread compared to other models. 
Manual listening evaluations of matched pairs from all models further confirm that the fine-tuned CLAP and PANNs consistently produce pairs that match with human auditory perception. 
In conclusion, we use the fine-tuned CLAP in our copy detection analysis in the main text.

\end{appendices}
\end{document}